\documentclass[conference]{IEEEtran}
\usepackage{booktabs} 
\usepackage{graphicx}
\usepackage{amsmath}
\usepackage{amsfonts}
\usepackage{caption}
\usepackage{subcaption}
\usepackage{algorithm}
\usepackage[noend]{algpseudocode}
\usepackage{float}
\usepackage{url}
\usepackage{siunitx}
\usepackage{multirow}
\graphicspath{{figs/}} 
\usepackage{etoolbox}
\usepackage{tikz}
\usepackage{slashbox}

\usepackage{times}
\usepackage{xcolor}
\usepackage{soul}
\usepackage[utf8]{inputenc}

\makeatletter
\newcommand{\algrule}[1][.2pt]{\par\vskip.5\baselineskip\hrule height #1\par\vskip.5\baselineskip}
\makeatother

\usepackage{algorithm}
\usepackage[noend]{algpseudocode}
\usepackage{xfrac}

\graphicspath{{figs/}} 
\newcommand{\sys}{DeepSigns}

\begin{document}
\title{DeepSigns: A Generic Watermarking Framework for \\ Protecting the Ownership of Deep Learning Models}

\author{
    Bita Darvish Rouhani, Huili Chen, and Farinaz Koushanfar\\
    University of California San Diego\\
    bita@ucsd.edu, huc044@eng.ucsd.edu, farinaz@ucsd.edu
} 

\maketitle
\begin{abstract}
Deep Learning (DL) models have caused a paradigm shift in our ability to comprehend raw data in various important fields, ranging from intelligence warfare and healthcare to autonomous transportation and automated manufacturing. A practical concern, in the rush to adopt DL models as a service, is protecting the models against Intellectual Property (IP) infringement. The DL models are commonly built by allocating significant computational resources that process vast amounts of proprietary training data. The resulting models are therefore considered to be the IP of the model builder and need to be protected to preserve the owner's competitive advantage. 

This paper proposes DeepSigns, a novel end-to-end IP protection framework that enables insertion of coherent digital watermarks in contemporary DL models. DeepSigns, for the first time, introduces a generic watermarking methodology that can be used for protecting DL owner's IP rights in both white-box and black-box settings, where the adversary may or may not have the knowledge of the model internals. The suggested methodology is based on embedding the owner's signature (watermark) in the probability density function (pdf) of the data abstraction obtained in different layers of a DL model. DeepSigns can demonstrably withstand various removal and transformation attacks, including model compression, model fine-tuning, and watermark overwriting. Proof-of-concept evaluations on MNIST, and CIFAR10 datasets, as well as a wide variety of neural network architectures including Wide Residual Networks, Convolution Neural Networks, and Multi-Layer Perceptrons corroborate DeepSigns' effectiveness and applicability. 
\end{abstract}

\section{Introduction}
The fourth industrial revolution empowered by machine learning algorithms is underway. The popular class of deep learning models and other contemporary machine learning methods are enabling this revolution by providing a significant leap in accuracy and functionality of the underlying model. Several applications are already undergoing serious transformative changes due to the integration of intelligence, including (but not limited to) social networks, autonomous transportation, automated manufacturing, natural language processing, intelligence warfare and smart health~\cite{lecun2015deep, deng2014deep, goodfellow2016deep, ribeiro2015mlaas}. 

Deep learning is an empirical field in which training a highly accurate model requires: (i) Having access to a massive collection of mostly labeled data that furnishes comprehensive coverage of potential scenarios that might appear in the target application. (ii) Allocating substantial computing resources to fine-tune the underlying model topology (i.e., type and number of hidden layers), hyper-parameters (i.e., learning rate, batch size, etc.), and DL weights in order to obtain the most accurate model. Given the costly process of designing and training a deep neural network, DL models are typically considered to be the intellectual property of the model builder. Protection of the models against IP infringement is particularly important for deep neural networks to preserve the competitive advantage of the DL model owner and ensure the receipt of continuous query requests by clients if the model is deployed in the cloud as a service. 

Embedding digital watermarks into deep neural networks is a key enabler for reliable technology transfer. A digital watermark is a type of marker covertly embedded in a signal or IP, including audio, videos, images, or functional designs. Digital watermarks are commonly adopted to identify ownership of the copyright of such a signal or function. Watermarking has been immensely leveraged over the past decade to protect the ownership of multimedia and video content, as well as functional artifacts such as digital integrated circuits~\cite{furht2004multimedia, hartung1999multimedia, qu2007intellectual,cox1997secure, lu2004multimedia}. Extension of watermarking techniques to deep learning models, however, is still in its infancy. 

DL models can be used in either a white-box or a black-box setting. In a white-box setting, the model parameters are public and shared with a third-party. Model sharing is a common approach in the machine learning field (e.g., the Model Zoo by Caffe Developers, and Alexa Skills by Amazon). Note that even though models are voluntarily shared with the public, it is important to protect pertinent IP and preserve the copyright of the original owner. In the black-box setting, the model details are not publicly shared and the model is only available to execute as a remote black-box Application Programming Interface (API). Most of the DL APIs deployed in cloud servers fall within the black-box category. 

Authors in~\cite{uchida2017embedding, nagai2018digital} propose a watermarking approach for embedding the IP information in the \textit{static} content of convolutional neural networks (i.e., weight matrices). Although this work provides a significant leap as the first attempt to watermark neural networks, it poses (at least) three limitations as we shall discuss in Section~\ref{sec:compprior}: (i) It incurs a bounded watermarking capacity due to the use of static properties of a model (weights) as opposed to using dynamic content (activations). Note that the weights of a neural network are invariable (static) during the execution phase, regardless of the data passing through the model. The activations, however, are dynamic and both \textit{data- and model-dependent}. As such, we argue that using activations (instead of static weights) provides more flexibility for watermarking purposes. (ii) It is not robust against overwriting the original embedded watermark by a third-party. (iii) It targets white-box settings and is inapplicable to black-box scenarios. 

\begin{table*}
\centering
\caption{Requirements for an effective watermarking of deep neural networks.}
\label{tab:required}
\scalebox{1}{
\begin{tabular}{|l||p{14.2cm}|}
\hline
\multicolumn{1}{|l||}{\textbf{Requirements}}   & \multicolumn{1}{|c|}{\textbf{Description}} \\ \hline \hline
\textbf{Fidelity}     & The functionality (e.g., accuracy) of the target neural network shall not be degraded as a result of watermark embedding. \\ \hline
\textbf{Reliability}  & The watermarking methodology shall yield minimal false negatives; the watermarked model shall be effectively detected using the pertinent keys. \\ \hline
\textbf{Robustness}   & The watermarking methodology shall be resilient against model modifications such as compression/pruning, fine-tuning, and/or watermark overwriting. \\ \hline
\textbf{Integrity }   & The watermarking methodology shall yield minimal false alarms (a.k.a., false positives); the watermarked model should be uniquely identified using the pertinent keys. \\ \hline
\textbf{Capacity }    & The watermarking methodology shall be capable of embedding a large amount of information in the target neural network while satisfying other requirements (e.g., fidelity, reliability, etc.).  \\ \hline
\textbf{Efficiency}   & The communication and computational overhead of watermark embedding and extraction/detection shall be negligible. \\ \hline
\textbf{Security }    & The watermark shall leave no tangible footprints in the target neural network; thus, an unauthorized individual cannot detect the presence of a watermark in the model. \\ \hline
\textbf{Generalizability} & The watermarking methodology shall be applicable in both white-box and black-box settings. \\ \hline
\end{tabular}}
\end{table*}

More recent studies in~\cite{merrer2017adversarial, yossi} propose 1-bit watermarking methodologies that are applicable to black-box models.\footnote{1-bit and 0-bit watermarking are used interchangeably in the literature.} These approaches are built upon model boundary modification and the use of random adversarial samples that lie near decision boundaries. Adversarial samples are known to be statistically unstable, meaning that the adversarial samples crafted for a model are not necessarily mis-classified by another network~\cite{ grosse2017statistical, rouhanisafe}. Therefore, even though the proposed approaches in~\cite{merrer2017adversarial, yossi} yield a high watermark detection rate (a.k.a. true positive rate), they are also too sensitive to hyper-parameter tuning and usually lead to a high \textit{false alarm rate}. Note that false ownership proofs based upon watermark extraction, in turn, jeopardize the integrity of the proposed watermarking methodology and render the use of watermarks for IP protection ineffective.

This paper proposes \sys{}, a novel end-to-end framework that empowers coherent integration of robust digital watermarks in contemporary deep learning models with no drop in overall prediction accuracy. The embedded watermarks can be triggered by a set of corresponding input keys to remotely detect the existence of the pertinent neural network in a third-party DL service. \sys{}, for the first time, introduces a \textit{generic} functional watermarking methodology that is applicable to both white-box and black-box settings. Unlike prior works that directly embed the watermark information in the static content (weights) of the pertinent model, \sys{} works by embedding an arbitrary N-bit string into the probability density function (pdf) of the activation sets in various layers of a deep neural network. The proposed methodology is simultaneously \textit{data- and model-dependent}, meaning that the watermark information is embedded in the dynamic content of the DL network and can only be triggered by passing specific input data to the model. Our suggested method leaves no visible impacts on the static properties of the DL model, such as the histogram of the weight matrices.

We provide a comprehensive set of quantitative and qualitative metrics that shall be evaluated to corroborate the effectiveness of current and pending watermarking methodologies for deep neural networks (Section~\ref{sec:requirements}). We demonstrate the robustness of our proposed framework with respect to state-of-the-art removal and transformative attacks, including model compression/pruning, model fine-tuning, and watermark overwriting. Extensive evaluation across various DL model topologies - including residual networks, convolutional neural networks, and multi-layer perceptrons - confirms the applicability of the proposed watermarking framework in different settings without requiring excessive hyper-parameter tuning to avoid false alarms and/or accuracy drop. The explicit contributions of this paper are as follows:
\begin{itemize}
    \item Proposing \sys{}, the first end-to-end framework for systematic deep learning IP protection that works in both white-box and black-box settings. A novel watermarking methodology is introduced to encode the pdf of the DL model and effectively trace the IP ownership. \sys{} is significantly more robust against removal and transformation attacks compared to prior works.
    
    \item Providing a comprehensive set of metrics to assess the performance of watermark embedding methods for DL models. These metrics enable effective quantitative and qualitative comparison of current and pending DL model protection methods that might be proposed in the future. 
    
    \item Devising an application programming interface to facilitate the adoption of \sys{} watermarking methodology for training various DL models, including convolutional, residual, and fully-connected networks. 
    
    \item Performing extensive proof-of-concept evaluations on various benchmarks. Our evaluations demonstrate \sys{}' effectiveness to protect the IP of an arbitrary DL model and establish the ownership of the model builder. 
\end{itemize}

\begin{figure*}[ht!]
\centering
\includegraphics[width=0.9\textwidth]{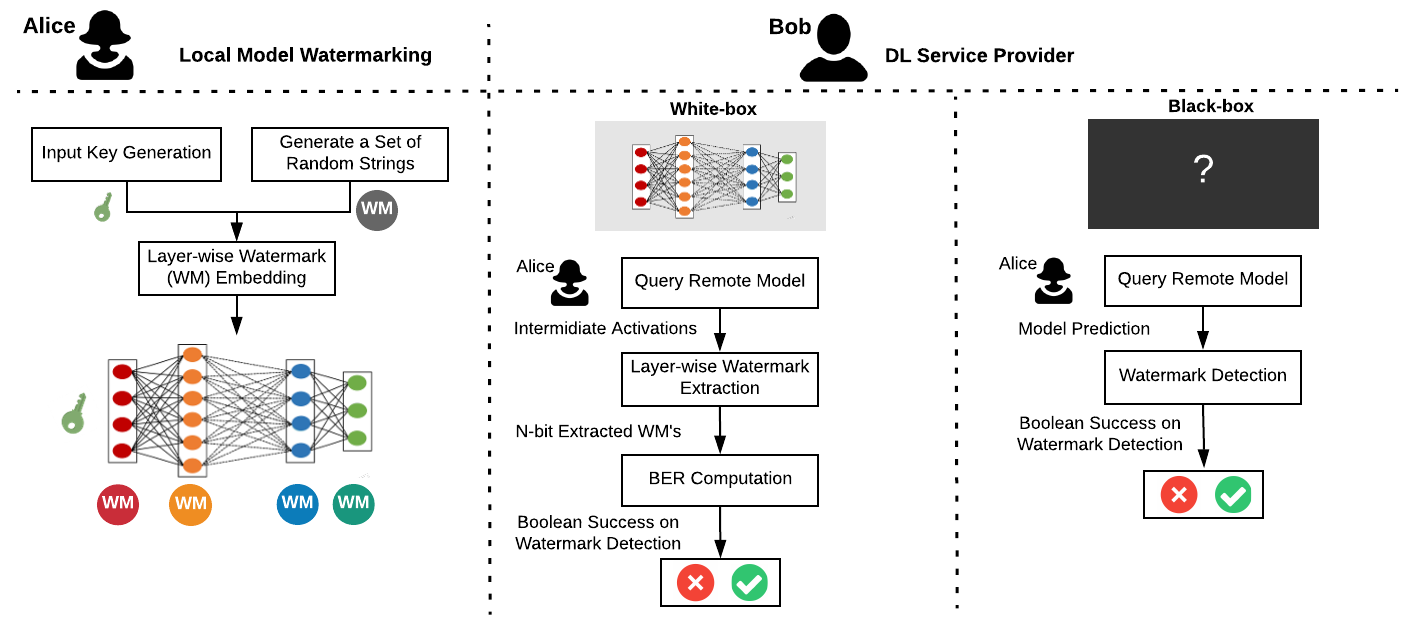} 
\caption{\label{fig:global} \sys{} Global Flow: \sys{} performs functional watermarking on DL models by simultaneously embedding a set of binary WM information in the pdf of the activation set acquired at each intermediate layer and the output layer. Typically, a specific set of inputs (keys) is used for extracting the embedded watermark. In our case, the inputs triggering the ingrained binary random strings are used as the key for the detection of IP infringement in both white-box and black-box settings.}
\end{figure*}

\section{Watermarking Requirements}{\label{sec:requirements}}
There are a set of minimal requirements that should be addressed to design a robust digital watermark. Table~\ref{tab:required} details the requirements for an effective watermarking methodology for DL models. In addition to previously suggested requirements in~\cite{uchida2017embedding, merrer2017adversarial}, we believe reliability, integrity, and generalizability are three other major factors that need to be considered when designing a practical DL watermarking methodology. 

Reliability is important because the embedded watermark should be accurately extracted using the pertinent keys; the model owner is thereby able to detect any misuse of her model with a high probability. Integrity ensures that the IP infringement detection policy yields a minimal number of false alarms, meaning that there is a very low chance of falsely proving the ownership of the model used by a third-party. Generalizability is another main factor in developing an effective watermarking methodology. Generalizability is particularly important since the model owner does not know beforehand whether her model will be misused in a black-box or white-box setting by a third-party. Nevertheless, the model owner should be able to detect IP infringement in both settings. \sys{} satisfies all the requirements listed in Table~\ref{tab:required} as shown by our experiments in Section~\ref{sec:eval}.

\vspace{0.5em}
\noindent\textbf{Potential Attack Scenarios.} To validate the robustness of a potential DL watermarking approach, one should evaluate the performance of the proposed methodology against (at least) three types of contemporary attacks: \textbf{(i) Model fine-tuning}. This type of attack involves re-training of the original model to alter the model parameters and find a new local minimum while preserving the accuracy. \textbf{(ii) Model pruning}. Model pruning is a commonly used approach for efficient execution of neural networks, particularly on embedded devices. We consider model pruning as another attack approach that might affect the watermark extraction/detection. \textbf{(iii) Watermark overwriting}. A third-party user who is aware of the methodology used to embed the watermark in the model (but not the owner's private watermark information) may try to embed a new watermark in the DL network and overwrite the original one. The objective of an overwriting attack is to insert an additional watermark in the model and render the original watermark unreadable. A watermarking methodology should be robust against fine-tuning, pruning, and overwriting for effective IP protection.

\section{Global Flow}{\label{sec:glob}}
Figure~\ref{fig:global} demonstrates the high-level block diagram of the \sys{} framework. To protect the IP of a particular neural network, the model owner (a.k.a. Alice) first must locally embed the watermark (WM) information into her neural network. Embedding the watermark involves three main steps: (i) Generating a set of N-bit binary random strings to be embedded in the pdf distribution of different layers in the target neural network. (ii) Creating specific input keys to later trigger the corresponding WM strings after watermark embedding. (iii) Training (fine-tuning) the neural network with particular constraints enforced by the WM information within intermediate activation maps of the target DL model. Details of each step are discussed in Section~\ref{sec:functW}. Note that local model watermarking is a one-time task only performed by the owner before model distribution.

Once the neural network is locally trained by Alice to include the pertinent watermark information, the model is ready to be deployed by a third-party DL service provider (a.k.a. Bob). Bob can either leverage Alice's model as a black-box API or a white-box model. To prove the ownership of the model, Alice queries the remote service provider using her specific input keys that she has initially selected to trigger the WM information. She then obtains the corresponding intermediate activations (in the white-box setting) or the final model prediction (in the black-box setting). The acquired activations/predictions are then used to extract the embedded watermarks and detect whether Alice's model is used by Bob within the underlying DL service or not. The details of watermark extraction are outlined in Section~\ref{sec:WE}.

\section{Functional Watermarking}{\label{sec:functW}}
Deep learning models possess non-convex loss surfaces with many local minima that are likely to yield an accuracy (on test data) very close to another approximate model~\cite{choromanska2015loss, rouhani2017deep3}. The \sys{} framework is built upon the fact that there is not a unique solution for modern non-convex optimization problems used in deep neural networks. \sys{} works by iteratively massaging the corresponding pdf of data abstractions to incorporate the desired watermarking information within each layer of the neural network. This watermarking information can later be used to claim ownership of the neural network or detect IP infringement.

In many real-world DL applications, the activation maps obtained in the intermediate (a.k.a. hidden) layers roughly follow a Gaussian distribution~\cite{ioffe2015batch, patel2015probabilistic, lin2016fixed}. In this paper, we consider a Gaussian Mixture Model (GMM) as the prior probability to characterize the data distribution at each hidden layer.\footnote{We emphasize that our proposed approach is rather generic and is not restricted to the GMM distribution; the GMM distribution can be replaced with any other prior distribution depending on the application.} The last layer (a.k.a. output layer) is an exception since the output can be a discrete variable (e.g., class label) in a large category of DL applications. As such, \sys{} governs the hidden (\textit{Section~\ref{sec:hidden}}) and output (\textit{Section~\ref{sec:output}}) layers differently.

\subsection{Watermarking Intermediate Layers}{\label{sec:hidden}}
To accommodate for our GMM prior distribution assumption, we suggest adding the following term to the conventional cross-entropy loss function ($loss_0$) used for training deep neural networks:\\
\begin{equation} \label{eq:opt}
\scalebox{0.99}{
$\lambda_1~(~\underbrace{\|\mu^l_{y^*} - f^l(x, \theta)\|_2^2 ~-~ \Sigma_{i\ne y^*} \|\mu^l_i - f^l(x, \theta)\|_2^2}_{\bf{loss_1}}~)$.}\\
\end{equation}\\
Here, $\lambda_1$ is a trade-off hyper-parameter that specifies the contribution of the additive loss term. The additive loss function ($loss_1$) aims to minimize the entanglement between data features (activations) belonging to different classes while decreasing the inner-class diversity. This loss function, in turn, helps to augment data features so that they approximately fit a GMM distribution. On one hand, a large value of $\lambda_1$ makes the data activations follow a strict GMM distribution, but large $\lambda_1$ values might also impact the final accuracy of the model due to limited contribution of the accuracy-specific cross-entropy loss function ($loss_0$). On the other hand, a very small value of $\lambda_1$ is not adequate to make the activations adhere to a GMM distribution. Note that the default distribution of activations may not strictly follow a GMM. We set the value $\lambda_1$ to $0.01$ in all our experiments. 

In Equation~(\ref{eq:opt}), $\theta$ is the set of model parameters (i.e., weights and biases), $f^l(x, \theta)$ is the activation map corresponding to input sample $x$ at the $l^{th}$ layer, $y^*$ is the ground-truth label, and $\mu^l_i$ denotes the mean value of the Gaussian distribution at layer $l$ that best fits the data abstractions belonging to class $i$. In \sys{} framework, the watermark information is embedded in the mean value of the pertinent Gaussian mixture distribution. The mean values $\mu^l_i$ and intermediate feature vectors $f^l(x, \theta)$ in Equation~\ref{eq:opt} are \textbf{trainable variables} that are iteratively learned and fine-tuned during the training process of the target deep neural network.

\vspace{0.5em}
\noindent \textbf{Watermark embedding.} To watermark the target neural network, the model owner (Alice) first needs to generate the designated WM information for each intermediate layer of her model. Algorithm 1 summarizes the process of watermark embedding for intermediate layers. In the following passages, we explicitly discuss each of the steps outlined in Algorithm 1. The model owner shall repeat Steps 1 through 3 for each layer that she wants to eventually watermark and sum up the corresponding loss functions for each layer in Step 4 to train the pertinent DL model.

\algnewcommand\algorithmicinput{\textbf{INPUT:}}
\algnewcommand\INPUT{\item[\algorithmicinput]}
\algnewcommand\algorithmicoutput{\textbf{OUTPUT:}}
\algnewcommand\OUTPUT{\item[\algorithmicoutput]}












\vspace{0.5em}
\noindent \textbf{\tikz\draw[black, fill=black] (0,0) circle (1.2ex) node[white] {1};} Choosing one (or more) random indices between $1$ and $S$ with no replacement: Each index corresponds to one of the Gaussian distributions in the target mixture model that contains a total of $S$ Gaussians. For classification tasks, we set the value $S$ equal to the number of classes in the target application. The mean values of the selected distributions $\mu_i^l$ are then used to carry the watermark information generated in Steps 2 and 3 as discussed below. 

\vspace{0.5em}
\noindent \textbf{\tikz\draw[black, fill=black] (0,0) circle (1.2ex) node[white] {2};}  Designating an arbitrary binary string to be embedded in the target model: The elements (a.k.a., bits) of the binary string are independently and identically distributed (i.i.d). Henceforth, we refer to this binary string as the vector $b\in \{0, 1\}^{s\times N}$ where $s$ is the number of selected distributions (Step 1) to carry the watermarking information, and $N$ is a owner-defined parameter indicating the desired length of the digital watermark embedded at the mean value of each selected Gaussian distribution.

\vspace{0.5em}
\noindent \textbf{\tikz\draw[black, fill=black] (0,0) circle (1.2ex) node[white] {3};}  Specifying a random projection matrix ($A$): The projection matrix is used to map the selected centers in Step 1 into the binary vector chosen in Step 2. The transformation is denoted as the following: \\
\begin{equation} \label{eq:trans}
\scalebox{0.99}{
$\begin{aligned}
G_\sigma^{s\times N}  &= Sigmoid~(\mu^{s\times M}~\textbf{.}~A^{M\times N}), \\
b^{s\times N}~&=~Hard\_Thresholding~(G_\sigma^{s\times N},~0.5). \\
\end{aligned}$}
\end{equation} 
\noindent Here, $M$ is the size of the feature space in the pertinent layer, and $\mu^{s\times M}$ denotes the concatenated mean values of the selected distributions. In our experiments, we use a standard normal distribution $\mathcal{N}(0, 1)$ to generate the WM projection matrix ($A$). Using i.i.d. samples drawn from a normal distribution ensures that each bit of the binary string is embedded into all the features associated with the selected centers (mean values). The $\sigma$ notation in Equation~\ref{eq:trans} is used as a subscript to indicate the deployment of the Sigmoid function. The output of Sigmoid has a value between 0 and 1. Given the random nature of the binary string, we decide to set the threshold in Equation~\ref{eq:trans} to $0.5$, which is the expected value of Sigmoid. The \textit{Hard\_Thresholding} function denoted in Equation~\ref{eq:trans} maps the values in $G_\sigma$ that are greater than $0.5$ to ones and the values less than $0.5$ to zeros. This threshold value can be easily changed in our API if the user decides to change it for their application. A value greater the $0.5$ means that the binary string has a higher probability to include more zeros than ones. This setting, in turn, is useful for users who use biased random number generators.

\vspace{0.5em}
\noindent \textbf{\tikz\draw[black, fill=black] (0,0) circle (1.2ex) node[white] {4};} Training the DL model to embed the pertinent watermark information: The process of computing the vector $G_\sigma$ is differentiable. Thereby, for a selected set of projection matrice ($A$) and binary strings ($b$), the selected centers (Gaussian mean values) can be adjusted/trained via back-propagation such that the Hamming distance between the binarized projected centers and the actual WM vectors $b$ is minimized (ideally zero). To do so, one needs to add the following term to the overall loss function for each specific layer of the underlying deep neural network:\\
\begin{equation} \label{eq:loss2}
\scalebox{0.99}{
$- \lambda_2 ~\underbrace {\sum\limits_{j=1}^{N} \sum\limits_{k=1}^{s} (b^{kj}~ \text{ln}(G_\sigma^{kj}) + (1-b^{kj})~\text{ln}(1-G_\sigma^{kj}))}_{\bf{loss_2}}.$ }\\
\end{equation}\\
Here, the variable $\lambda_2$ is a hyper-parameter that determines the contribution of $loss_2$ in the process of training the neural network. All three loss functions ($loss_0$, $loss_1$, and $loss_2$) are simultaneously used to train/fine-tuned the underlying neural network. We used Stochastic Gradient Descent (SGD) in all our experiments to optimize the DL model parameters with the explicit constraints outlined in Equations~\ref{eq:opt}~and~\ref{eq:loss2}. This optimization aims to find the best distribution of the activation sets by iterative data subspace alignment in order to obtain the highest accuracy while embedding the WM vectors. We set the $\lambda_2$ variable to $0.01$ in all our experiments, unless mentioned otherwise. As shown in Section~\ref{sec:eval}, our method is robust on various benchmarks even though the hyper-parameters are not explicitly tuned for each application.

\begin{figure}[ht!]
\centering
\includegraphics[width=0.46\textwidth]{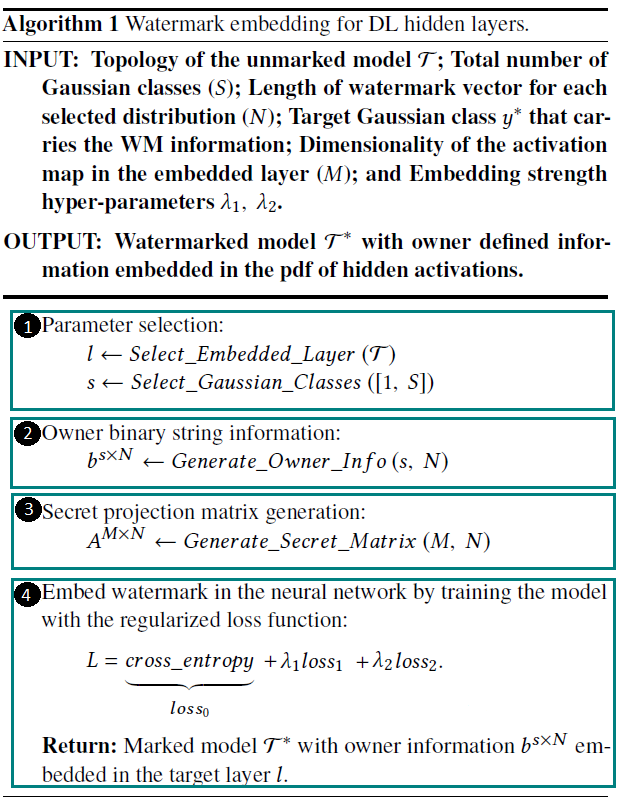} 
\vspace{-1em}
\end{figure}

\subsection{Watermarking The Output Layer}
\begin{figure}[ht!]
\centering
\includegraphics[width=0.45\textwidth]{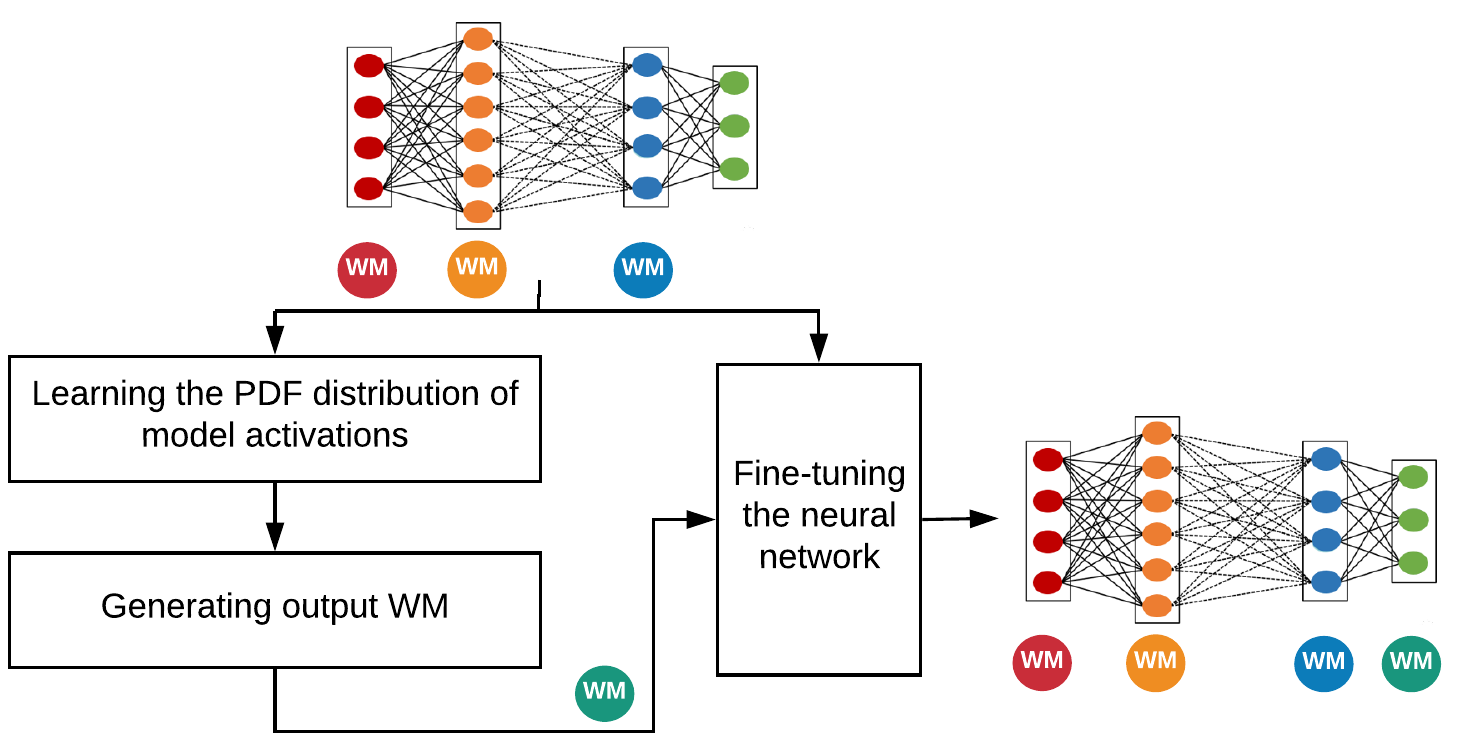} 
\caption{\label{fig:Woutput} High-level overview of watermarking the output layer in a neural network. Output watermarking is a post-processing step performed after embedding the selected binary WMs in the intermediate (hidden) layers.}
\end{figure}

{\label{sec:output}}
Neural network prediction in the very last layer of a DL model needs to closely match the ground-truth data (e.g., training labels in a classification task) in order to have the maximum possible accuracy. As such, instead of directly regularizing the activation set of the output layer, we choose to adjust the tails of the decision boundaries to incorporate a desired statistical bias in the network as a 1-bit watermark. We focus, in particular, on classification tasks using deep neural networks. Watermarking the output layer is a post-processing step that shall be performed after training the model as discussed in Section~\ref{sec:hidden}. Figure~\ref{fig:Woutput} illustrates the high-level block-diagram of the \sys{} framework used for watermarking the output layer of the underlying DL model.

\noindent The workflow of embedding watermark in the output layer is summarized in Algorithm~2. In the following, we explicitly discuss each of the steps outlined in Algorithm 2. 

\begin{figure}[ht!]
\centering
\includegraphics[width=0.33\textwidth]{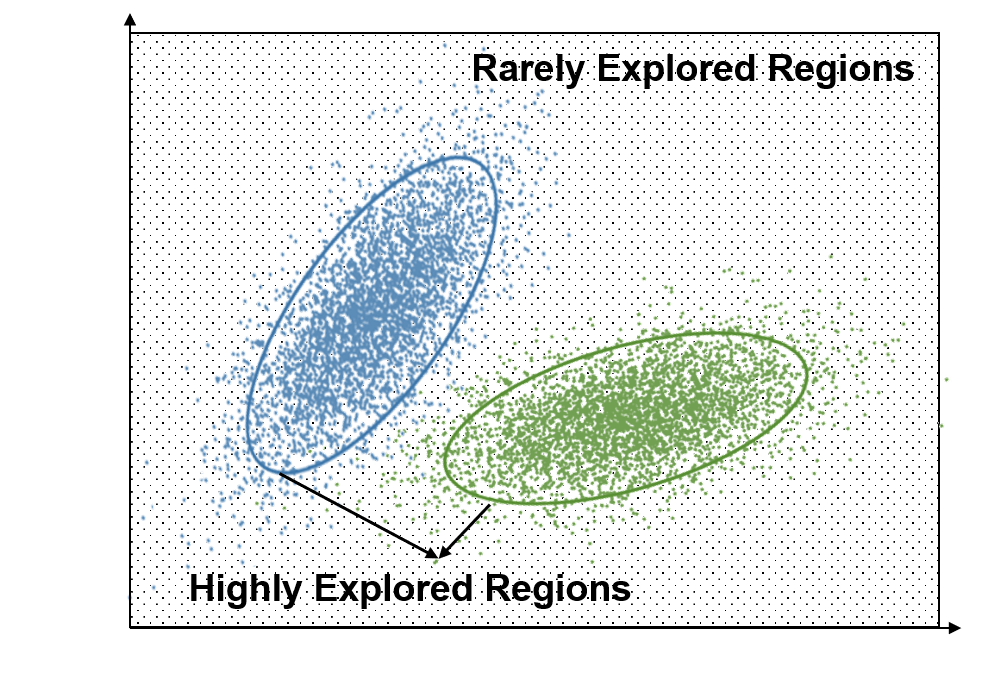} 
\caption{\label{fig:explored} Due to the high dimensionality of deep learning models and limited access to labeled training data (the blue and green dots in the figure), there are sub-spaces within the DL model that are rarely explored. \sys{} exploits this mainly unused capacity to embed the watermark information while minimally affecting ultimate accuracy.}
\end{figure}

\vspace{0.3em}
\noindent {\tikz\draw[black,fill=black] (-1.2em,-1.2em) rectangle (-0.2em,-0.2em) node[pos=.5, white] {1};} Learning the pdf distribution of the activations in each intermediate layer as discussed in Section~\ref{sec:hidden}: The acquired probability density function, in turn, gives us an insight into both the regions of latent space that are thoroughly occupied by the training data and the regions that are only covered by the tail of the GMM distribution, which we refer to as rarely explored regions. Figure~\ref{fig:explored} illustrates a simple example of two clustered Gaussian distribution spreading in a two-dimensional subspace.

\vspace{0.3em}
\noindent {\tikz\draw[black,fill=black] (-1.2em,-1.2em) rectangle (-0.2em,-0.2em) node[pos=.5, white] {2};} Generating a set of $K$ unique random input samples to be used as the watermarking keys in step 3: Each selected random sample should be passed through the pre-trained neural network in order to make sure its latent features lie within the unused regions (Step 1). If the number of training data within a $\epsilon$-ball of the random sample is fewer than a threshold, we accept that sample as one of the watermark keys. Otherwise, a new random sample is generated to replace the previous sample. A corresponding random ground-truth vector is generated and assigned to each selected input key sample. For instance, in a classification application, each random input is associated with a randomly selected class.

On one hand, it is desirable to have a high watermark detection rate after WM embedding. On the other hand, one needs to ensure a low false positive rate to address the integrity requirement. As such, we start off by setting the initial key size to be larger than the owner's desired value $K^{'} > K$ and generate $\left\{ X^{key'}, Y^{key'} \right\}$ accordingly. The target model is then fine-tuned (Step 3) using a mixture of the generated keys and a subset of original training data. After fine-tuning, only keys that are simultaneously correctly classified by the marked model and incorrectly predicted by the unmarked model are appropriate candidates that satisfy both a high detection rate and a low false positive (Step 4). In our experiments, we set $K^{'}=20\times K$ where $K$ is the desired key length selected by the model owner (Alice). 

\vspace{0.3em}
\noindent {\tikz\draw[black,fill=black] (-1.2em,-1.2em) rectangle (-0.2em,-0.2em) node[pos=.5, white] {3};} Fine-tuning the pre-trained neural network with the selected random watermarks in Step 2: The model shall be retrained such that the neural network has exact predictions (e.g., an accuracy greater than 99\%) for selected key samples. In our experiments, we use the same optimizer setting originally used for training the neural network, except that the learning rate is reduced by a factor of 10 to prevent accuracy drop in the prediction of legitimate input data.

\vspace{0.3em}
\noindent {\tikz\draw[black,fill=black] (-1.2em,-1.2em) rectangle (-0.2em,-0.2em) node[pos=.5, white] {4};} Selecting the final input key set to trigger the embedded watermark in the output layer: To do so, we first find out the indices of initial input keys that are correctly classified by the marked model. Next, we identify the indices of input key samples that are not classified correctly by the original DL model before fine-tuning in Step 3. The common indices between these two sets are proper candidates to be considered as the final key. A random subset of applicable input key samples is then selected based on the required key size ($K$) that is defined by the model owner.

\begin{figure}[ht!]
\centering
\includegraphics[width=0.46\textwidth]{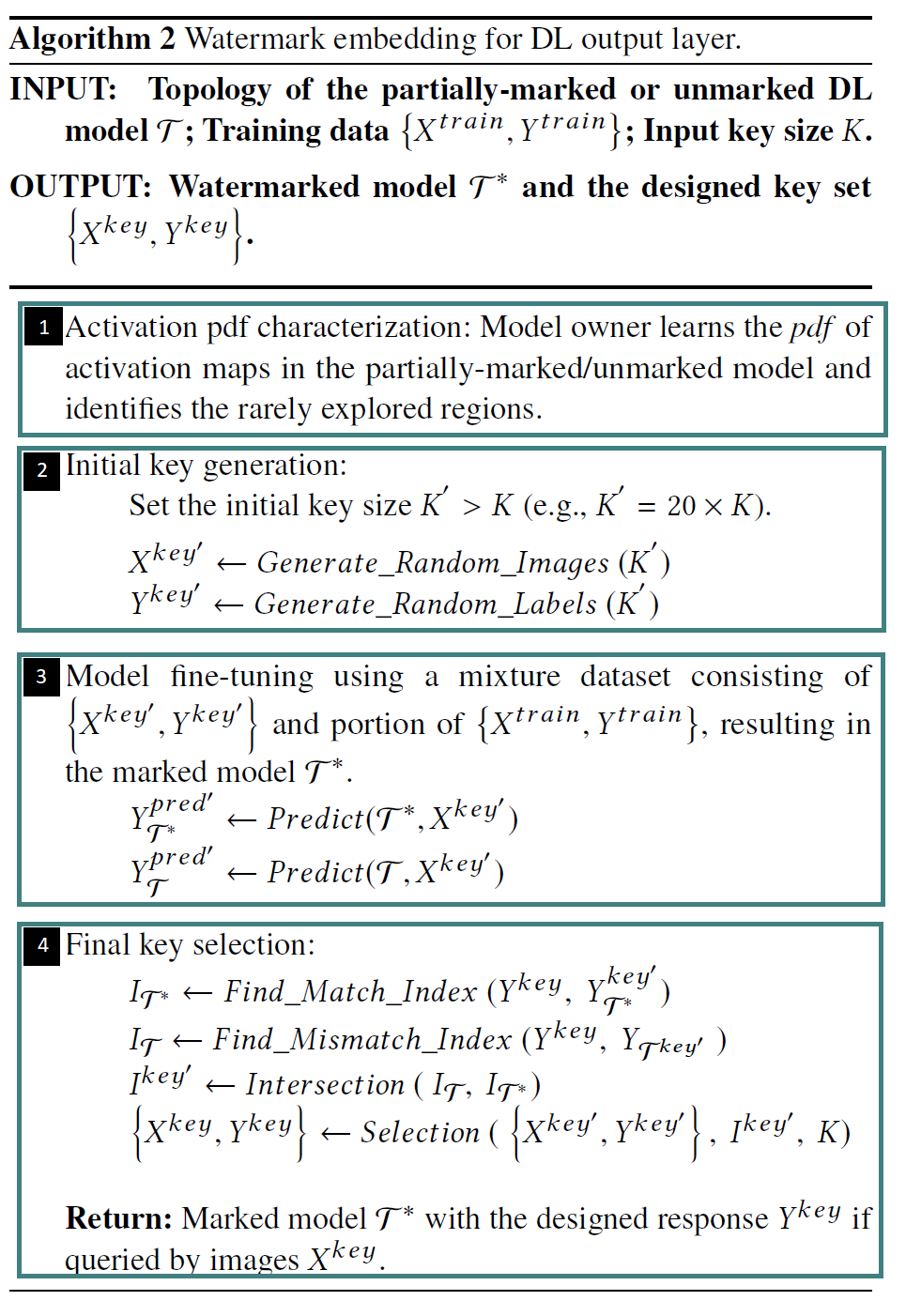} 
\vspace{-2em}
\end{figure}

It is worth noting that the watermark information embedded in the output layer can be extracted even in settings where the DL model is used as a black-box API by a third-party (Bob). Recently a method called frontier stitching is proposed to perform 1-bit watermarking in black-box settings~\cite{merrer2017adversarial}. Our proposed approach is different in the sense that we use random samples that lie within the tail regions of the probability density function spanned by the model, as opposed to relying on adversarial samples that lie close to the boundaries~\cite{goodfellow2014explaining, papernot2017practical, grosse2017statistical, rouhanisafe}. Adversarial samples are known to be statistically unstable, meaning that the adversarial samples carefully crafted for a model are not necessarily mis-classified by another network. As shown previously in~\cite{merrer2017adversarial}, frontier stitching is highly vulnerable to the hyper-parameter selection of the watermarking detection policy and may lead to a high false positive if it is not precisely tuned; thus jeopardizing the integrity requirement. As we empirically verify in Section~\ref{sec:eval}, \sys{} overcomes this integrity concern by selecting random samples within the unused space of the model. This is due to the fact that the unused regions in the space spanned by a model are specific to that model, whereas the decision boundaries for a given task are often highly correlated among various models.

\section{Watermarking Extraction}{\label{sec:WE}}
For watermark extraction, the model owner (Alice) needs to send a set of queries to the DL service provider (Bob). The queries include the input keys discussed in Sections~\ref{sec:hidden} and \ref{sec:output}. In the case of black-box usage of a neural network, Alice can only retrieve model predictions for the queried samples, whereas in the white-box setting, the intermediate activations can also be recovered. In the rest of this section, we explicitly discuss the decision policy for IP infringement detection in both white-box and black-box scenarios. Given the high dimensionality of neural networks, the probability of collision for two honest model owners is very unlikely. Two honest owners should have the exact same weight initialization, projection matrix, binary string, and input keys to end up with the same watermark. In case of a malicious user, the misdetection probability is evaluated for the various attacks discussed in Section~\ref{sec:eval}. 

\subsection{Decision Policy: White-box Setting}
To extract watermark information from intermediate (hidden) layers, Alice must follow five main steps. Algorithm 3 outlines the decision policy for the white-box scenarios. 

\vspace{0.3em}
\noindent \textbf{(I)} Submitting queries to the remote DL service provider using the selected input keys. To do so, Alice first collects a subset of the input training data belonging to the selected watermarked classes ($y^*$ in Algorithm 1). In our experiments, we use a subset of 1\% of training data as the input key. \textbf{(II)} Acquiring the activation features corresponding to the input keys. \textbf{(III)} Computing statistical mean value of the activation features obtained by passing the selected input keys in Step I. The acquired mean values are used as an approximation of the Gaussian centers that are supposed to carry the watermark information. \textbf{(IV)} Using the mean values obtained in Step III and her private projection matrix $A$ to extract the pertinent binary string following the protocol outlined in Equation~\ref{eq:trans}. \textbf{(V)} Measuring the Bit Error Rate (BER) between the original watermark string and the extracted string from Step IV. Note that in case of a mismatch between Alice and Bob models, a random watermark will be extracted which, in turn, yields a very high BER.  

\begin{algorithm}[ht!]
\caption*{\textbf{Algorithm 3} Watermark extraction in white-box setting. }
\label{alg:WM_extract_whitebox}
\begin{algorithmic}[1]

\INPUT \textbf{Remote DL model $\mathcal{T}^{'}$; Target Gaussian class $y^{*}$ that carries the WM information; Location of the embedded layer $l$; Training data $\left\{ X^{train}, Y^{train} \right\}$; Owner's WM information $b$.}

\vspace{0.5em}
\OUTPUT \textbf{Extracted watermark $b^{'}$ and BER. }
\algrule[1pt]
\vspace{0.2em}
\State Key set generation: 

$\left\{ X^{key},~Y^{key} \right\} \gets Select\_Pairs(\left\{ X^{train},~ Y^{train} \right\},~y^{*})$

\vspace{0.5em}
\State Acquire activation features: 

$f^{l}(x,~\theta) \gets Forward\_Pass~(\mathcal{T^{'}},~X^{key},~l)$

\vspace{0.5em}

\State Compute mean of activation:

$\mu^{s \times M} \gets Compute\_Mean~( f^{l}(x,~\theta) )$

\vspace{0.5em}

\State Extract WM:

$G_{\sigma}^{s \times N } \gets Sigmoid~(\mu^{s\times M}~\textbf{.}~A^{M\times N})$

\vspace{0.5em}

$b^{'} \gets Hard\_Thresholding(G_{\sigma}^{s \times N },~0.5) $

\State Evaluate BER:

$BER \gets Number\_of\_Bit\_Mismatches~(b,~b^{'})$

\noindent \textbf{Return:} BER of the queried DL model. 
 
\end{algorithmic}
\end{algorithm}

\vspace{0.5em}
\noindent \textbf{Computation and communication overheads.} From Bob's point of view, the computation cost is equivalent to the cost of one forward pass in the pertinent DL model for each query from Alice. From Alice's point of view, the computation cost is divided into two terms. The first term is proportional to $\mathcal{O}(M)$ to compute the statistical mean in Step 3 outlined in the Algorithm 3. Here, $M$ denotes the feature space size in the target hidden layer. The second term corresponds to the computation of matrix multiplication in Step 4 of Algorithm 3, which incurs a cost of $\mathcal{O}(MN)$. The communication cost for Bob is equivalent to the input key length multiplied by the feature size of intermediate activations ($M$), and the communication cost for Alice is the input key size multiplied by the input feature size for each sample.  

\vspace{-0.5em}
\subsection{Decision Policy: Black-box Setting}{\label{subsec:bdp}}

To verify the presence of the watermark in the output layer, Alice needs to statistically analyze Bob's responses to a set of input keys. To do so, she must follow four main steps: \textbf{(I)} Submitting queries to the remote DL service provider using the randomly selected input keys ($X^{key}$) as discussed in Section~\ref{sec:output}. \textbf{(II)} Acquiring the output labels corresponding to the input keys. \textbf{(III)} Computing the number of mismatches between the model predictions and Alice's ground-truth labels. \textbf{(IV)} Thresholding the number of mismatches to derive the final decision. If the number of mismatches is less than a threshold, it means that the model used by Bob possesses a high similarity to the network owned by Alice. Otherwise, the two models are not replicas. When the two models are the exact duplicate of one another, the number of mismatches will be zero and Alice can safely claim the ownership of the neural network used by the third-party.

\begin{algorithm}[ht!]
\caption*{ \textbf{Algorithm 4} Watermark detection in the black-box setting. }
\label{alg:WM_extract_blackbox}
\begin{algorithmic}[1]

\INPUT \textbf{ Remote DL model $\mathcal{T}^{'}$; Owner's input key set $\left\{ X^{key}, Y^{key} \right\}$; Maximum tolerated number of mismatches $N_K$}

\vspace{0.5em}
\OUTPUT \textbf{One bit indicating the presence of the owner's WM in the remote DL model. }

\algrule[1pt]
\vspace{0.2em}

\State Alice sends her input keys $X^{key}$ to Bob $\mathcal{T}$.

\vspace{0.5em}
\State Inference by the remote model:

$Y^{pred} \gets Predict~(\mathcal{T^{'}},~X^{key})$

\vspace{0.5em}

\State Response comparison:

$ n_k \gets Count\_Mismatch~(Y^{pred},~Y^{key})$

\vspace{0.5em}

\State Decision making:

$Presence = 1~if~n_k~<~N_k~else~0$

\noindent \textbf{Return:} WM presence indicator ($Presence$)
 
\end{algorithmic}
\end{algorithm}

\begin{table*}
\centering
\caption{ Benchmark neural network architectures. Here, ${64C3(1)}$ indicates a convolutional layer with $64$ output channels and ${3\times3}$ filters applied with a stride of 2, ${MP2(1)}$ denotes a max-pooling layer over regions of size ${2\times2}$ and stride of 1, and ${512FC}$ is a fully-connected layer with $512$ output neurons. ReLU is used as the activation function in all benchmarks. }
\label{tab:bench}
\scalebox{0.92}{
\begin{tabular}{|c||c|cc||c|c|} \hline

\textbf{Dataset} & \textbf{Baseline Accuracy} & \multicolumn{2}{c||}{\textbf{Accuracy of Marked Model}} & \textbf{DL Model Type} & \textbf{DL Model Architecture} \\ \hline \hline

\multirow{2}{*}{MNIST}   &  \multirow{2}{*}{98.54\%}  & \multicolumn{1}{c|}{K~=~20}  & N~=~4  & \multirow{2}{*}{MLP} & \multirow{2}{*}{784-512FC-512FC-10FC}  \\ \cline{3-4} &    & \multicolumn{1}{c|}{98.59\%}  & 98.13\%   &  & \\ \hline

\multirow{2}{*}{CIFAR10} &  \multirow{2}{*}{78.47\%}  & \multicolumn{1}{c|}{K~=~20} & N~=~4 & \multirow{2}{*}{CNN} & 3*32*32-32C3(1)-32C3(1)-MP2(1) \\ \cline{3-4} &   &  \multicolumn{1}{c|}{81.46\%}  & 80.7\% & & -64C3(1)-64C3(1)-MP2(1)-512FC-10FC\\ \hline
                         
\multirow{2}{*}{CIFAR10} &  \multirow{2}{*}{91.42\%}                                                      & \multicolumn{1}{c|}{K~=~20}   & N~=~128 & \multirow{2}{*}{WideResNet} & \multirow{2}{*}{Please refer to \cite{zagoruyko2016wide}.} \\ \cline{3-4}&                             &  \multicolumn{1}{c|}{91.48\%} & 92.02\% &  &\\ \hline

\end{tabular}
} 
\end{table*}

In real-world settings, the target DL model might be slightly modified by Bob in both malicious or non-malicious ways. Examples of such modifications are model fine-tuning, model pruning, or WM overwriting. As such, the threshold used for WM detection should be greater than zero to withstand DL model modifications. The probability of a network (not owned by Alice) to make at least $n_k$ correct decision according to the Alice private keys is as follows:
\begin{equation} \label{eq:threshold}
\scalebox{1.05}{
$P(N_k > n_k| \mathcal{O}) = 1 - \sum\limits_{k=0}^{n_k} \binom{K}{k} (\frac{1}{C})^{K - k} (1 - \frac{1}{C})^{k}$,}
\end{equation}
where $\mathcal{O}$ is the oracle DL model used by Bob, $N_k$ is a random variable indicating the number of matched predictions of the two models compared against one another, $K$ is the input key length according to Section~\ref{sec:output}, and $C$ is the number of classes in the pertinent deep learning application. 

Algorithm 4 summarizes the decision policy for the black-box scenarios. Throughout our experiments, we use the decision policy $P(N_k > n_k| \mathcal{O}) > (1-1e^{-3})$ for watermark detection, where $P(N_k > n_k| \mathcal{O})$ is defined in Equation~\ref{eq:threshold}. The threshold value used in Step 5 of Algorithm 4 determines the trade-off between a low false positive rate and a high detection rate. As we empirically corroborate in Section~\ref{sec:eval}, \sys{} satisfies the reliability, integrity, and robustness requirements in various benchmarks and attack scenarios without demanding that the model owner explicitly fine-tune her decision policy hyper-parameters (e.g., the threshold in Equation~\ref{eq:threshold}).

\vspace{0.5em}
\noindent \textbf{Computation and communication overheads.} For Bob, the computation cost is equivalent to the cost of one forward pass through the underlying neural network per queried input key. For Alice, the computation cost is the cost of performing a simple counting to measure the number of mismatches between the output labels by Bob and the actual labels owned by Alice. The communication cost for Alice is equivalent to the key length multiplied by the size of the input layer in the target neural network. From Bob's point of view, the communication cost to send back the corresponding predicted output is the key length multiplied by the output layer size.

\section{Evaluations}{\label{sec:eval}}
We evaluate the performance of the \sys{} framework on various datasets including MNIST~\cite{lecun1998mnist} and CIFAR10~\cite{krizhevsky2009learning} with three different neural network architectures. Table~\ref{tab:bench} summarizes the neural network topologies used in each benchmark. In Table~\ref{tab:bench}, $K$ denotes the key size for watermarking the output layer and $N$ is the length of the WM used for watermarking the hidden layers. In all white-box related experiments, we use the second-to-last layer for watermarking. However, \sys{} is generic and the extension of watermarking multiple layers is also supported by our framework. In the block-box scenario, we use the very last layer for watermark embedding/detection. To facilitate watermark embedding and extraction in various DL models, we provide an accompanying TensorFlow-based~\cite{abadi2016tensorflow} API in which model owners can easily define their specific model topology, watermark information, training data, and pertinent hyper-parameters including decision policies' thresholds, WM embedding strength ($\lambda_1,~\lambda_2$), and selected layers to carry the WM information.

\begin{figure*}[ht!]
       \centering
\begin{subfigure}[b]{0.6\columnwidth}
    \includegraphics[width=\textwidth]{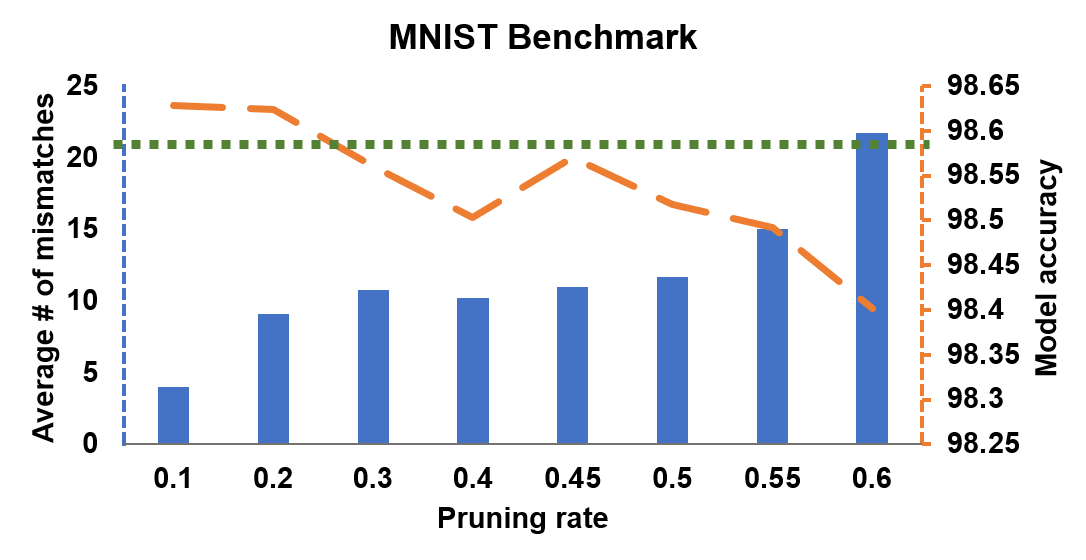}
    \caption{}
\end{subfigure}
\begin{subfigure}[b]{0.6\columnwidth}
    \includegraphics[width=\textwidth]{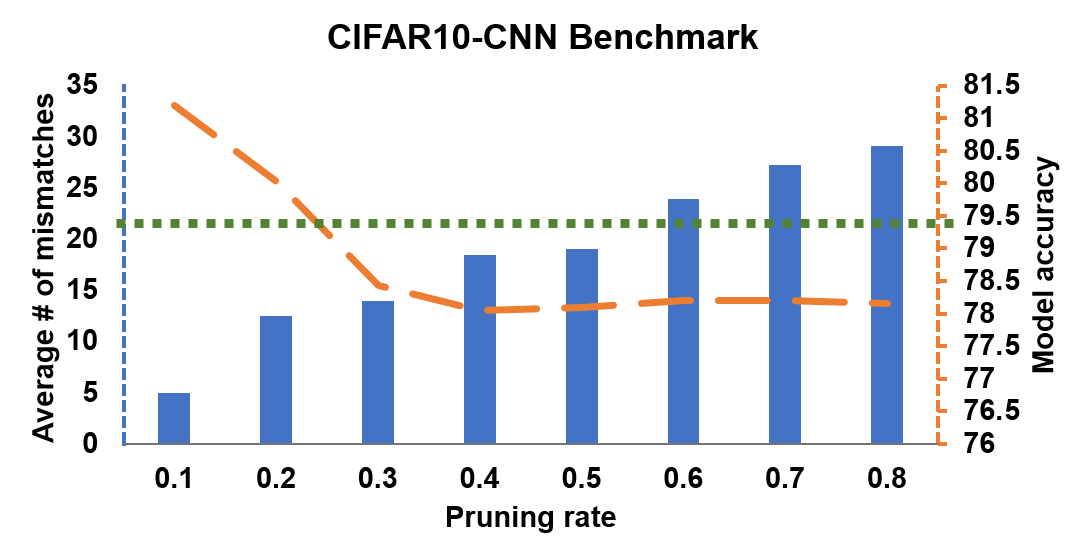}
    \caption{}
\end{subfigure}
\begin{subfigure}[b]{0.6\columnwidth}
    \includegraphics[width=\textwidth]{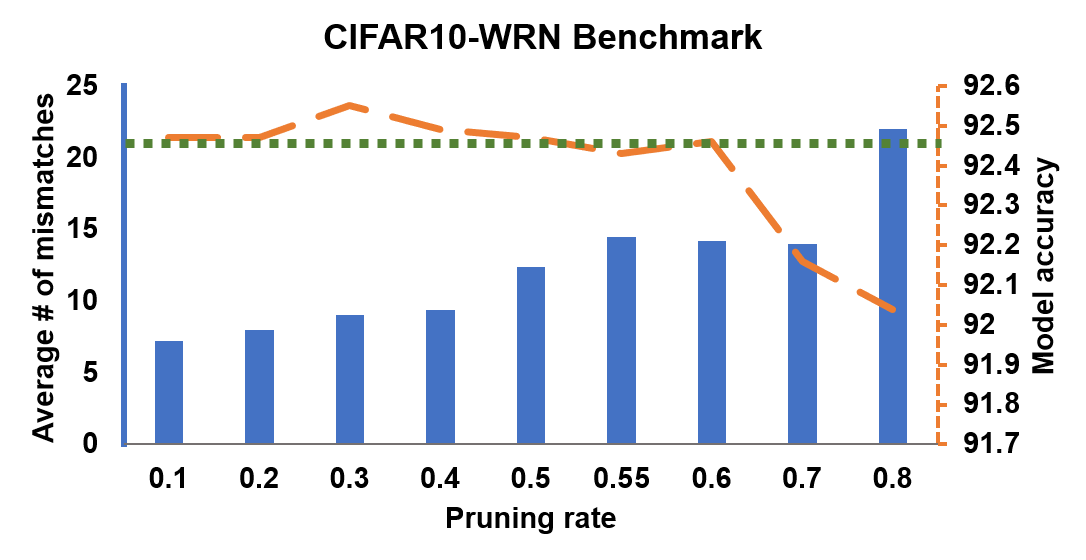}
    \caption{}
\end{subfigure}

\begin{subfigure}[b]{0.61\columnwidth}
    \includegraphics[width=\textwidth]{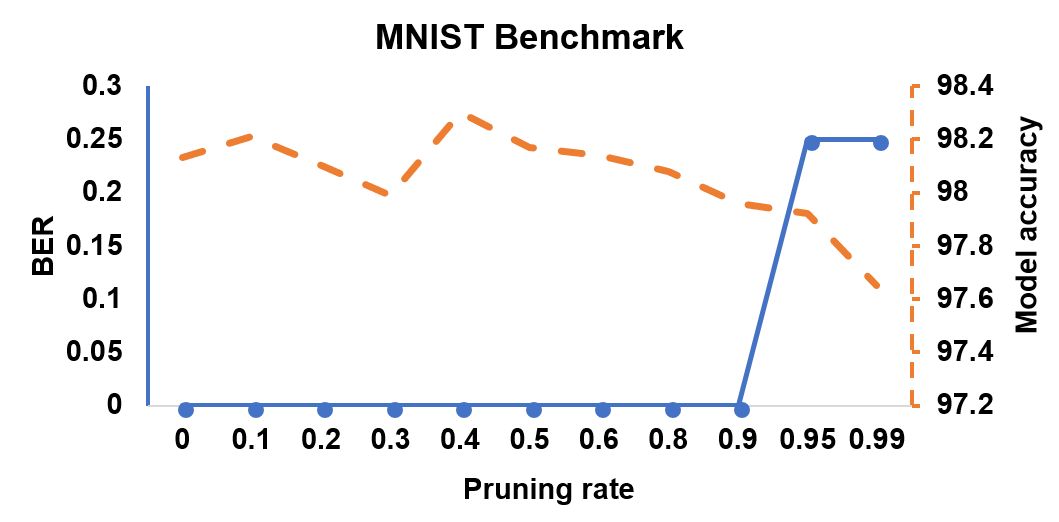}
    \caption{}
\end{subfigure}
\begin{subfigure}[b]{0.61\columnwidth}
    \includegraphics[width=\textwidth]{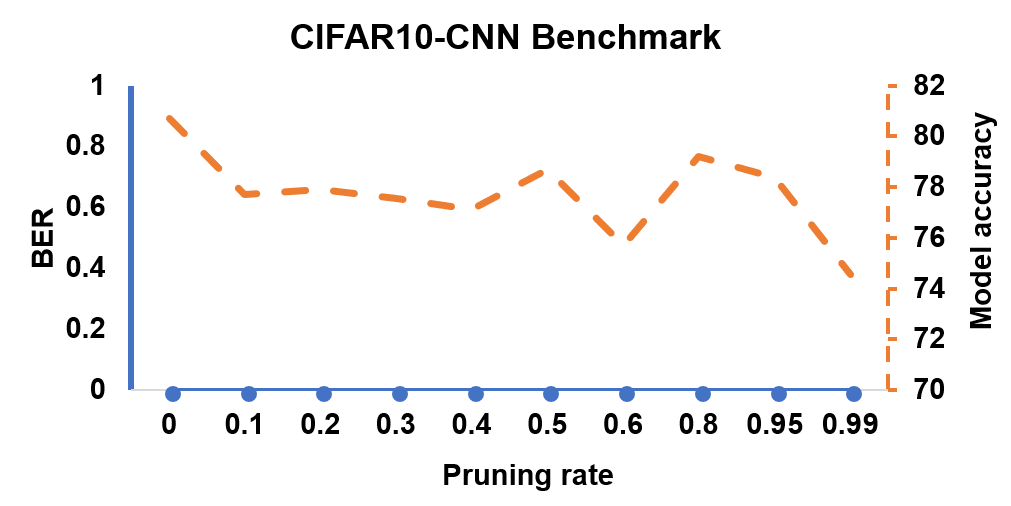}
    \caption{}
\end{subfigure}
\begin{subfigure}[b]{0.61\columnwidth}
    \includegraphics[width=\textwidth]{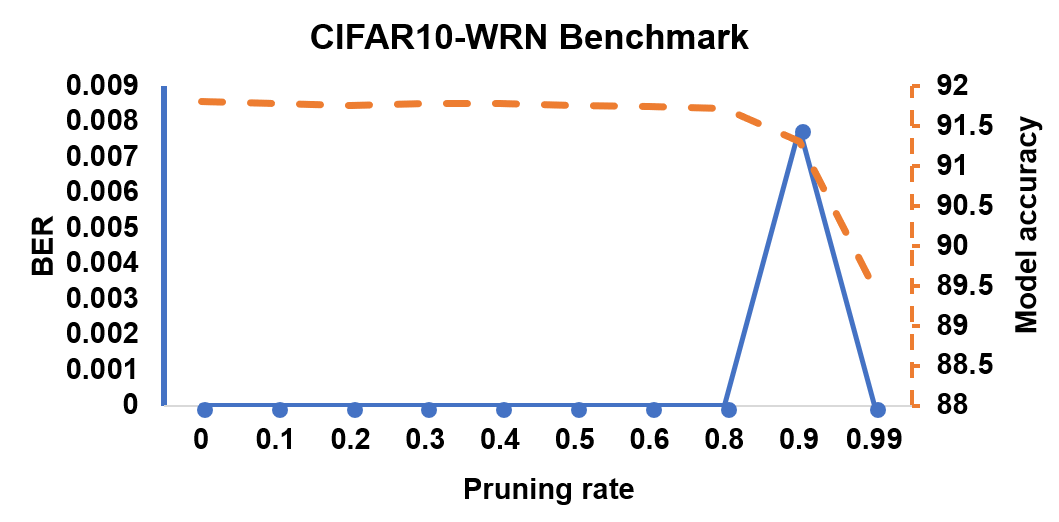}
    \caption{}
\end{subfigure}
\caption{\label{fig:pruning} Evaluation of the watermark's robustness against parameter pruning. Figures (a) through (c) (first row) illustrate for each of the benchmarks listed in Table~\ref{tab:bench} in the black-box setting. The horizontal green dotted line is the mismatch threshold obtained from Equation~(\ref{eq:threshold}). The orange dashed lines show the corresponding test accuracy for each pruning rate. Figures (d) through (f) (second row) show the results for the MNIST and CIFAR10 benchmarks in the white-box setting. The dashed lines demonstrate the pertinent accuracy per pruning rate.}
\end{figure*}

\begin{table*}
\centering
\caption{Robustness of the \sys{} framework against the model fine-tuning attack. The reported BER and the detection rate value are averaged over 10 different runs. A value of 1 in the last row of the table indicates that the embedded watermark is successfully detected, whereas a value of 0 indicates a false negative. For fine-tuning attacks, the WM-specific loss terms proposed in Section~\ref{sec:functW} are removed from the loss function and the model is retrained using the final learning rate of the original DL model. After fine-tuning, the DL model will converge to another local minimum that is not necessarily a better one (in terms of accuracy) for some benchmarks.}
\label{tab:fine_tuning}
\scalebox{0.8}{
\begin{tabular}{|l||ccc|ccc|ccc||ccc|ccc|ccc|}
\hline
\multirow{2}{*}{\textbf{Metrics}} &
\multicolumn{9}{c||}{\textbf{White-box}}  & \multicolumn{9}{c|}{\textbf{Black-box}} \\ \cline{2-19} 

& \multicolumn{3}{c|}{\textbf{MNIST}} & \multicolumn{3}{c|}{\textbf{CIFAR10-CNN}} & \multicolumn{3}{c||}{\textbf{CIFAR10-WRN}} &  \multicolumn{3}{c|}{\textbf{MNIST}} & \multicolumn{3}{c|}{\textbf{CIFAR10-CNN}} & \multicolumn{3}{c|}{\textbf{CIFAR10-WRN}} \\ \hline \hline

\textbf{Number of epochs} & 50 & 100 & 200  & 50 & 100 & 200  & 50 & 100 & 200  &    50 & 100 & 200  &    50 & 100 & 200 &    50 & 100 & 200     \\ \hline

\textbf{Accuracy} & 98.21    & 98.20    & 98.18    & 70.11     & 62.74     & 59.86     & 91.79 & 91.74 &  91.8 &  98.57 & 98.57 & 98.59 & 98.61    & 98.63    & 98.60   &   87.65     & 89.74     & 88.35     \\ \hline

\textbf{BER} & 0     & 0        & 0        & 0         & 0         & 0   & 0     & 0        & 0             & -        & -        & -        & -         & -         & -   & -        & -        & -             \\ \hline

\textbf{Detection success} & \textbf{1} & \textbf{1} & \textbf{1} & \textbf{1} & \textbf{1} & \textbf{1} & \textbf{1} & \textbf{1} & \textbf{1} & \textbf{1} & \textbf{1} & \textbf{1} & \textbf{1} & \textbf{1} & \textbf{1} & \textbf{1} & \textbf{1} & \textbf{1} \\ \hline 
\end{tabular}}
\end{table*}

\vspace{0.3em}
\noindent \sys{ } satisfies all the requirements listed in Table~\ref{tab:required}. In the rest of this section, we explicitly evaluate \sys{}' performance with respect to each requirement on three common DL benchmarks.

\subsection {Fidelity} 
The accuracy of the target neural network shall not be degraded after embedding the watermark information. Matching the accuracy of the unmarked model is referred to as \textit{fidelity}. Table~\ref{tab:bench} summarizes the baseline DL model accuracy (Column 2) and the accuracy of marked models (Column 3) after embedding the WM information. As demonstrated, \sys{} respects the fidelity requirement by simultaneously optimizing for the accuracy of the underlying model (e.g., cross-entropy loss function), as well as the additive WM-specific loss functions ($loss_1$ and $loss_2$) as discussed in Section~\ref{sec:functW}. In some cases (e.g. wide-ResNet benchmark), we even observe a slight accuracy improvement compared to the baseline. This improvement is mainly due to the fact that the additive loss functions (Equations~\ref{eq:opt} and~\ref{eq:loss2}) and/or exploiting rarely observed regions act as a form of a regularizer during the training phase of the target DL model. Regularization, in turn, helps the model to avoid over-fitting by inducing a small amount of noise to the DL model~\cite{goodfellow2016deep}.

\subsection {Reliability and Robustness} 
We evaluate the robustness of the \sys{} framework against three contemporary removal attacks as discussed in Section~\ref{sec:glob}. The potential attacks include parameter pruning~\cite{han2015learning, han2015deep, rouhani2016delight}, model fine-tuning~\cite{simonyan2014very, tajbakhsh2016convolutional}, and watermark overwriting~\cite{uchida2017embedding, johnson2001information}.

\vspace{0.7em}
\noindent \textbf{Parameter pruning.} We use the pruning approach proposed in~\cite{han2015learning} to compress the neural network. For pruning each layer of a neural network, we first set $\alpha\%$ of the parameters that possess the smallest weight values to zero. The obtained mask is then used to sparsely fine-tune the model to compensate for the accuracy drop induced by pruning using conventional cross-entropy loss function ($loss_0$). Figure~\ref{fig:pruning} illustrates the impact of pruning on watermark extraction/detection in both black-box and white-box settings. In the black-box experiments, \sys{} can tolerate up to $60\%$ and $35\%$ parameter pruning for the MNIST and CIFAR10 benchmarks, respectively, and up to $90\%$ and $99\%$ in the white-box related experiments. The WM lengths to perform watermarking in each benchmark are listed in Table~\ref{tab:bench}.

As demonstrated in Figure~\ref{fig:pruning}, in occasions where pruning the neural network yields a substantial bit error rate (BER) value, we observe that the sparse model suffers from a large accuracy loss compared to the baseline. As such, one cannot remove the embedded watermark in a neural network by excessive pruning of the parameters while attaining a comparable accuracy with the baseline.

\vspace{0.7em}
\noindent \textbf{Model fine-tuning.} Fine-tuning is another form of transformation attack that a third-party user might use to remove WM information. To perform this type of attack, one needs to retrain the target model using the original training data with the conventional cross-entropy loss function (excluding the watermarking specific loss functions). Table~\ref{tab:fine_tuning} summarizes the impact of fine-tuning on the watermark detection rate across all three benchmarks. 

There is a trade-off between the model accuracy and the success rate of watermark removal. If a third-party tries to fine-tune the DL model using a high learning rate with the goal of disrupting the underlying pdf of the activation maps and eventually remove WMs, he will face a large degradation in model accuracy. In our experiments, we use the same learning rate as the one in the final stage of DL training to perform the model fine-tuning attack. As demonstrated in Table~\ref{tab:fine_tuning}, \sys{} can successfully detect the watermark information even after fine-tuning the deep neural network for many epochs. Note that fine-tuning deep learning models makes the underlying neural network converge to another local minimum that is not necessarily equivalent to the original one in terms of the ultimate prediction accuracy.

\vspace{0.5em}
\noindent \textbf{Watermark overwriting.} Assuming the attacker is aware of the watermarking technique, he may attempt to damage the original watermark by embedding a new WM in the DL model. In practice, the attacker does not have any knowledge about the location of the watermarked layers. However, in our experiments, we consider the worst-case scenario in which the attacker knows where the WM is embedded but does not know the original watermark information. To perform the overwriting attack, the attacker follows the protocol discussed in Section~\ref{sec:functW} to embed a new set of watermark information (using a different projection matrix, binary vector, and input keys). Table~\ref{tab:overwrite} summarizes the results of watermark overwriting for all three benchmarks in the black-box setting. As shown, \sys{} is robust against the overwriting attack and can successfully detect the original embedded WM in the overwritten model. The decision thresholds shown in Table~\ref{tab:overwrite} for different key lengths are computed based on Equation~\ref{eq:threshold} as discussed in Section~\ref{subsec:bdp}. A bit error rate of zero is also observed in the white-box setting for all the three benchmarks after the overwriting attack. This further confirms the reliability and robustness of the \sys{}' watermarking approach against malicious attacks.

\begin{table}[ht!]
\centering
\caption{\sys{} is robust against overwriting attacks. In this experiment, the reported number of mismatches is the average value of multiple runs of the overwriting attack for the same model using different input key set. Since the average number of mismatches is smaller than the decision threshold (Equation~\ref{eq:threshold}) for each key length, \sys{} can successfully detect the original WM after the overwriting attack.}
\label{tab:overwrite}
\resizebox{0.48\textwidth}{!}{%
\begin{tabular}{|c||c|c||c|c||c|c||c|}
\hline
\multirow{2}{*}{} & \multicolumn{2}{c||}{\textbf{Average \# of mismatches}} & \multicolumn{2}{c||}{\textbf{Decision} \textbf{threshold}} & \multirow{2}{*}{\begin{tabular}[c]{@{}c@{}}\textbf{Detection} \\ \textbf{success}\end{tabular}} \\ \cline{2-5}
 & K = 20 & K = 30 & K = 20 & K = 30 &  \\ \hline \hline
\textbf{MNIST} & 8.3 & 15.4 & 13 & 21 & \textbf{1} \\ \hline
\textbf{CIFAR10-CNN} & 9.2 & 16.7 & 13 & 21 & \textbf{1} \\ \hline
\textbf{CIFAR10-WRN }& 8.5 & 10.2 & 13 & 21 & \textbf{1 }\\ \hline
\end{tabular}%
}
\end{table}

\begin{figure*}[ht!]
       \centering
\begin{subfigure}[b]{0.66\columnwidth}
    \includegraphics[width=\textwidth]{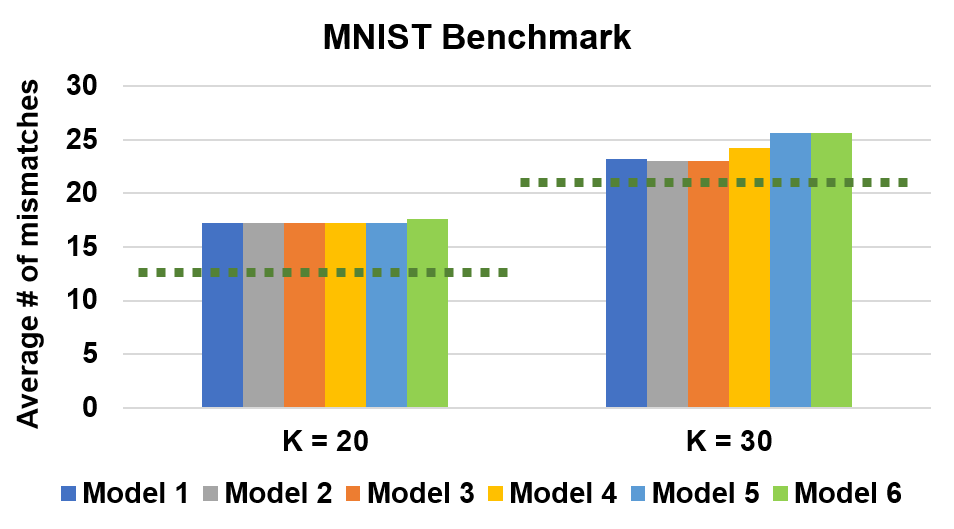}
    \caption{}
\end{subfigure}
\begin{subfigure}[b]{0.66\columnwidth}
    \includegraphics[width=\textwidth]{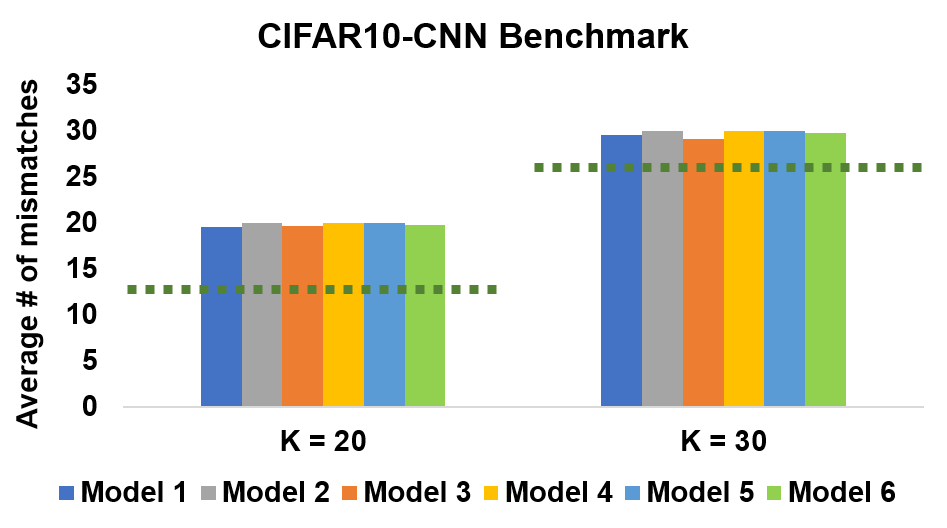}
    \caption{}
\end{subfigure}
\begin{subfigure}[b]{0.66\columnwidth}
    \includegraphics[width=\textwidth]{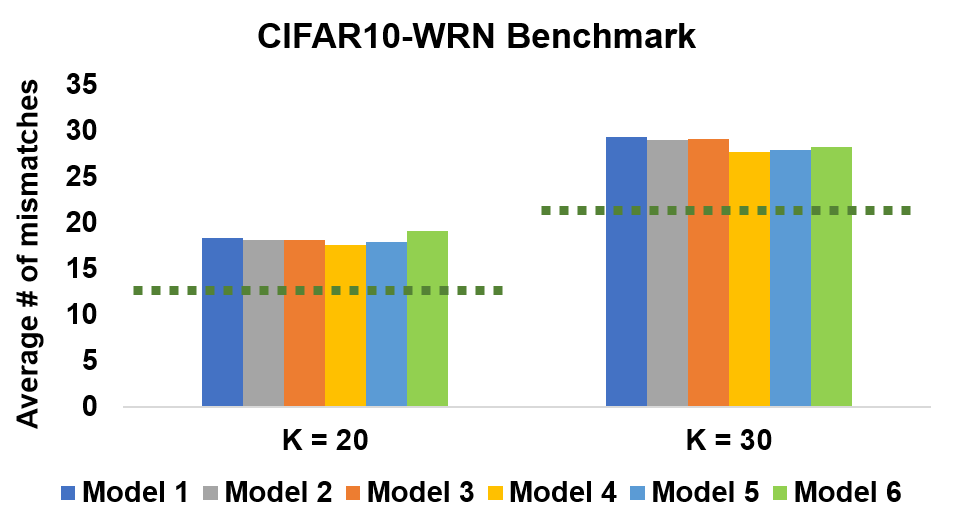}
    \caption{}
\end{subfigure}
\caption{\label{fig:BlackIntegrity} Integrity analysis of different benchmarks. The green dotted horizontal lines indicate the detection threshold for various WM lengths. The first three models (model 1-3) are neural networks with the same topology but different parameters compared with the marked model. The last three models (model 4-6) are neural networks with different topologies (~\cite{springenberg2014striving}, ~\cite{liang2015recurrent}, ~\cite{zagoruyko2016wide}). }
\end{figure*}

\begin{figure*}[ht!]
       \centering
       
\begin{subfigure}[b]{0.6\columnwidth}
    \includegraphics[width=\textwidth]{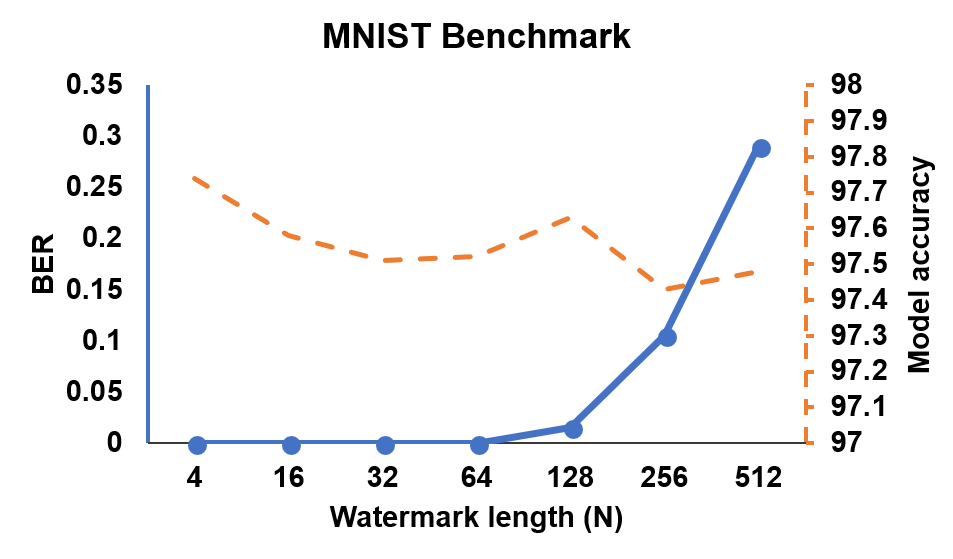}
    \caption{}
\end{subfigure}
\begin{subfigure}[b]{0.6\columnwidth}
    \includegraphics[width=\textwidth]{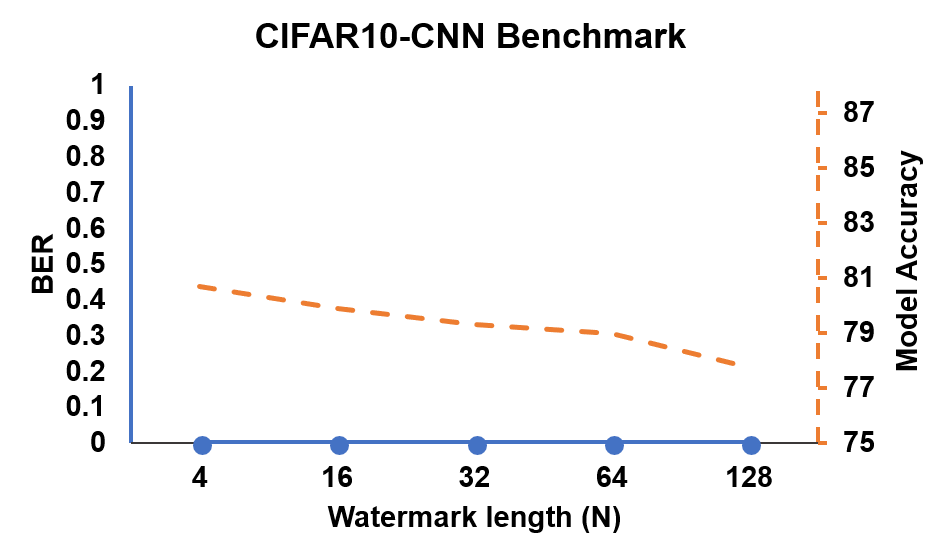}
    \caption{}
\end{subfigure}
\begin{subfigure}[b]{0.61\columnwidth}
    \includegraphics[width=\textwidth]{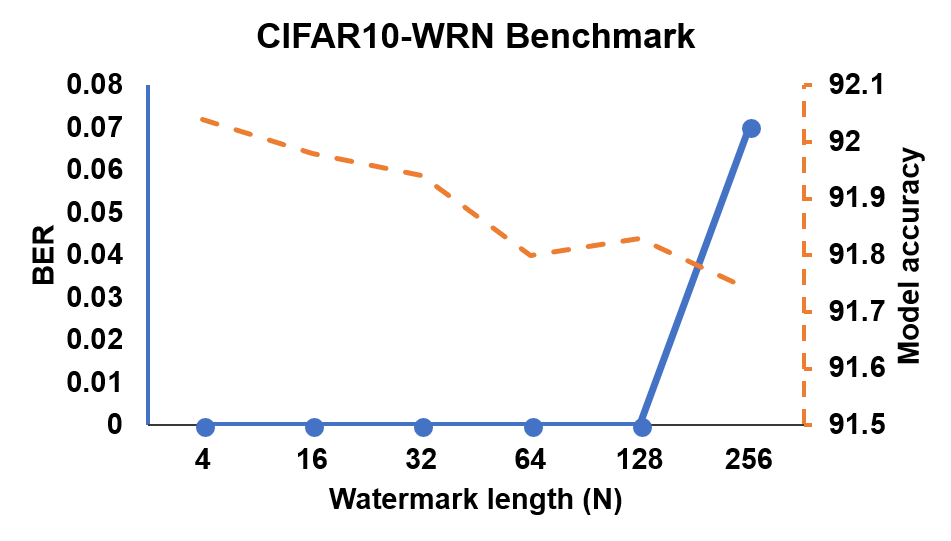}
    \caption{}
\end{subfigure}

\caption{\label{fig:BERCap} 
There is a trade-off between the length of WM vector (capacity) and watermark detection bit error rate. As the number of the embedded bits ($N$) increases, the test accuracy of the marked model decreases and the BER of the extracted WM increases. The trend indicates that embedding excessive amount of information in the WM impairs the fidelity and reliability.}
\end{figure*}

\subsection{Integrity}  Figure~\ref{fig:BlackIntegrity} illustrates the results of integrity evaluation in the black-box setting where unmarked models with the same (models 1 to 3) and different (models 4 to 6) topologies are queried by Alice's keys. As shown in Figure~\ref{fig:BlackIntegrity}, \sys{} satisfies the integrity criterion and has no false positives, which means the ownership of unmarked models will not be falsely proved. Note that unlike the black-box setting, in the white-box scenario, different topologies can be distinguished by one-to-one comparison of the architectures belonging to Alice and Bob. For the unmarked model with the same topology in the white-box setting, the integrity analysis is equivalent to model fine-tuning, for which the results are summarized in Table~\ref{tab:fine_tuning}.

\subsection{Capacity and Efficiency} The capacity of the white-box activation watermarking is assessed by embedding binary strings of different lengths in the intermediate layers. As shown in Figure~\ref{fig:BERCap}, \sys{} allows up to 64, 128, and 128 bits capacity for MNIST, CIFAR10-CNN, and CIFAR10-WRN benchmarks, respectively. Note that there is a trade-off between the capacity and accuracy which can be used by the IP owner (Alice) to embed a larger watermark in her neural network model if desired. For IP protection purposes, capacity is not an impediment criterion as long as the capacity is sufficient to contain the necessary WM information ($N~>~1$). Nevertheless, we have included this property in Table~\ref{tab:required} to have a comprehensive list of requirements.

In scenarios where the accuracy might be jeopardized due to excessive regularization of the intermediate activation features (e.g., when using a large watermarking strength for $\lambda_1$ and $\lambda_2$, a large WM length $N$, or in cases where the DL model is so compact that there are few free variables to carry the WM information), one can mitigate the accuracy drop by expanding each layer to include more free variables. Although the probability of finding a poor local minimum is non-zero for small-size networks, this probability decreases quickly with the expansion of network size~\cite{choromanska2015loss, dauphin2014identifying}, resulting in the accuracy compensation. Note that, in all our experiments, we do not expand the network to meet the baseline accuracy. The layer expansion technique is simply a suggestion for data scientists and DL model designers who are working on developing new architectures that might require a higher WM embedding capacity.

The computation and communication overhead in the \sys{} framework is a function of the network topology (i.e., the number of parameters/weights in the pertinent DL model), the selected input key length ($K$), and the size of the desired embedded watermark $N$. In Section~\ref{sec:WE}, we provide a detailed discussion on the computation and communication overhead of \sys{} in both white-box and black-box settings for watermark extraction. Note that watermark embedding is performed locally by Alice; therefore, there is no communication overhead for WM embedding. In all our experiments, training the DL models with WM-specific loss functions takes the same number of epochs compared to training the unmarked model (with only conventional cross-entropy loss) to obtain a certain accuracy. As such, there is no tangible extra computation overhead for WM embedding.

\subsection{Security}
\begin{figure}[ht!]
    \centering
\begin{subfigure}[b]{0.48\columnwidth}
    \includegraphics[width=\textwidth]{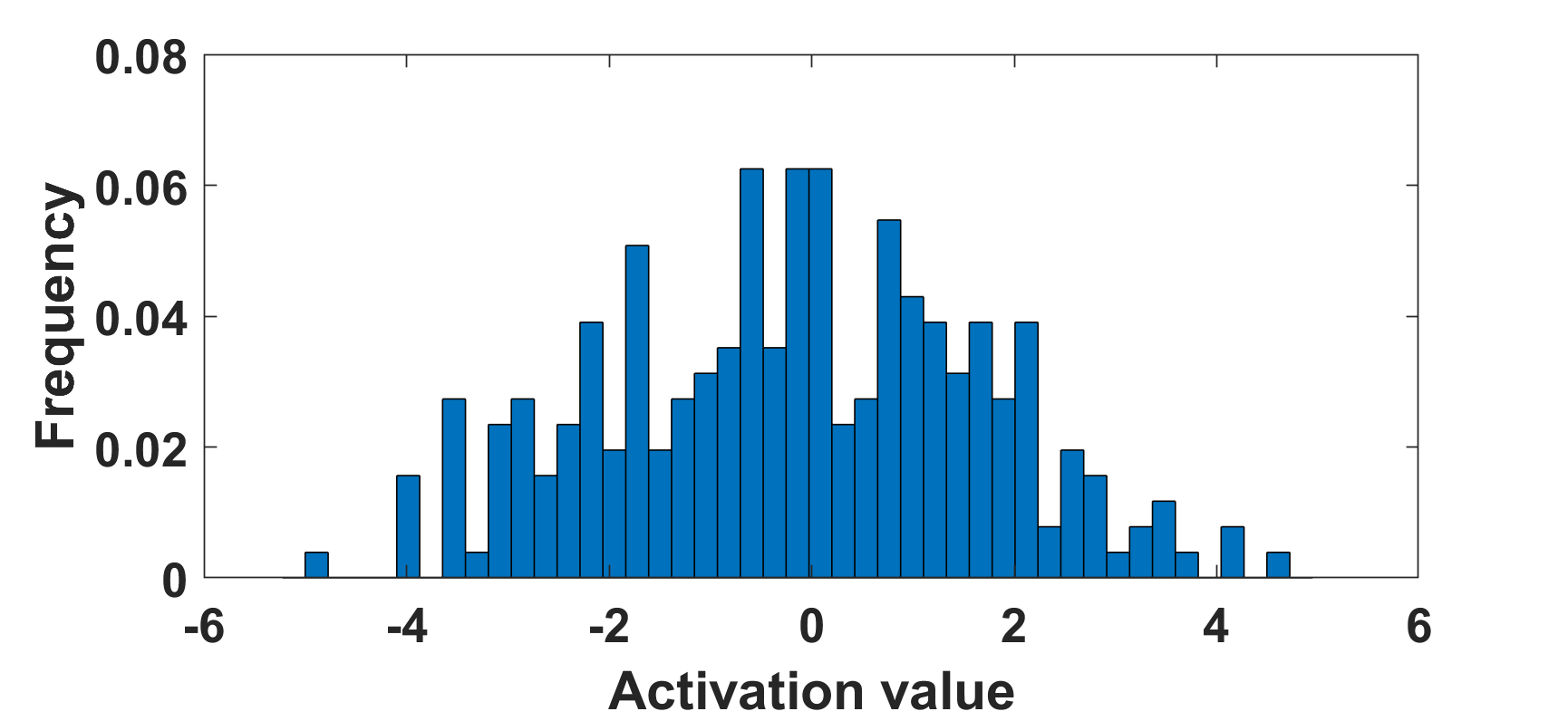}
    \caption{}
\end{subfigure}
\begin{subfigure}[b]{0.48\columnwidth}
    \includegraphics[width=\textwidth]{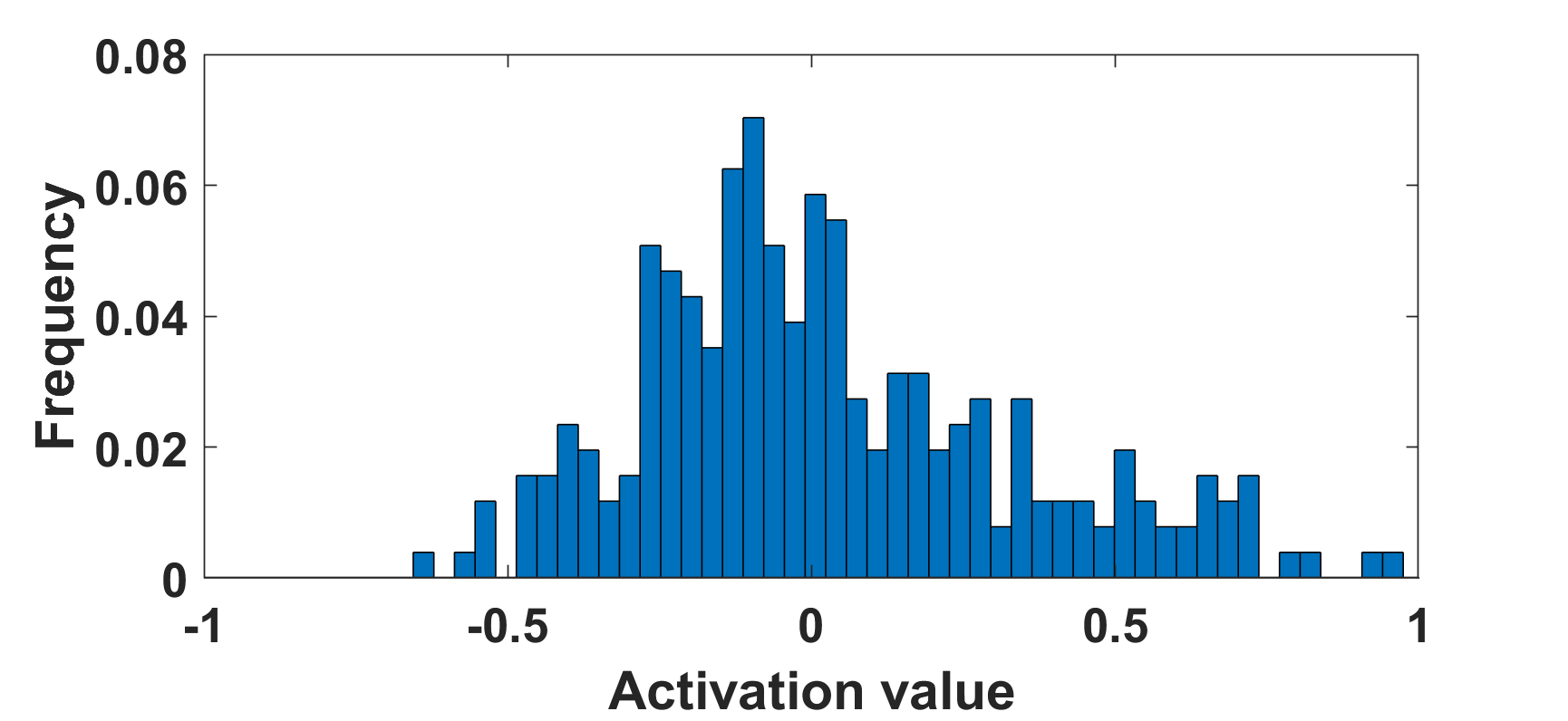}
    \caption{}
\end{subfigure}
\caption{\label{fig:Security} Distribution of the activation maps for marked (Figure a) and unmarked (Figure b) models. \sys{} preserves the intrinsic distribution spanned by the model while robustly embedding WM information. Note that the range of activations is not deterministic in different models and cannot be used by malicious users to detect the existence of a watermark.}
\end{figure}

As mentioned in Table~\ref{tab:required}, the embedding of the watermark should not leave noticeable changes in the probability distribution spanned by the target neural network. \sys{} satisfies the security requirement by preserving the intrinsic distribution of weights/activations. For instance, Figure~\ref{fig:Security} illustrates the histogram of the activations in the embedded layer in the marked model and the same layer in the unmarked model for the CIFAR10-WRN benchmark. 

\section{Comparison With Prior-Art}
\begin{figure}[ht!]
\includegraphics[width=0.48\textwidth]{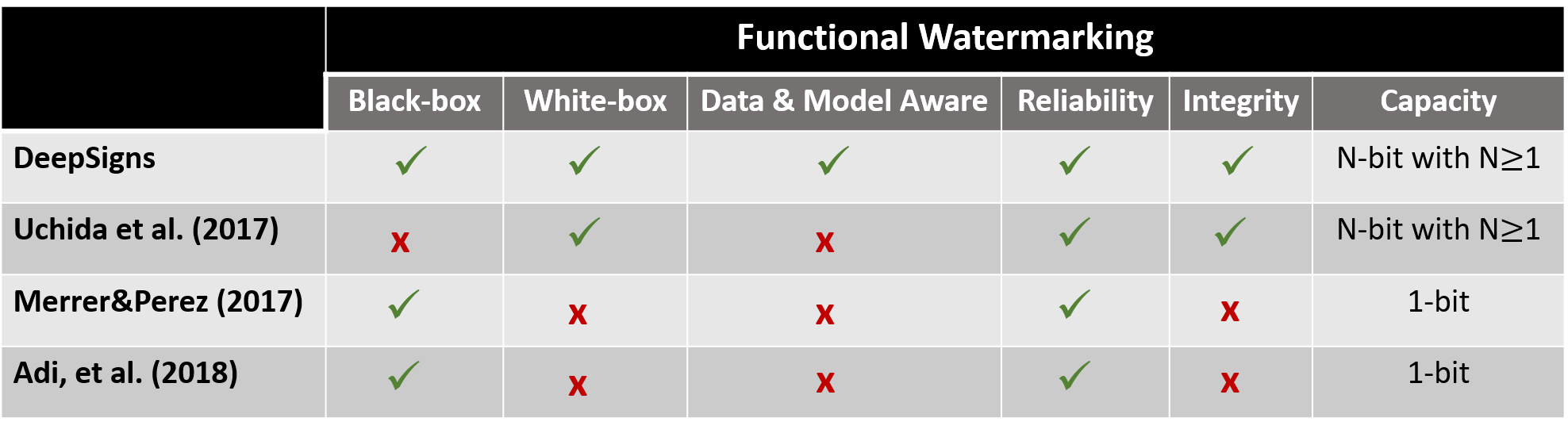}
\caption{\label{fig:comp} High-level comparison with prior-art deep learning watermarking frameworks.}
\end{figure}

{\label{sec:compprior}}
Figure~\ref{fig:comp} provides a high-level overview of general capabilities of existing DL WM frameworks. \sys{} satisfies the generalizability criterion and is applicable to both white-box and black-box settings. One may speculate that white-box and black-box scenarios can be treated equivalently by embedding the watermark information only in the output layer. In other words, one can simply ignore the white-box related information (intermediate layers) and simply treat the white-box setting as another black-box setting. We would like to emphasize that the output layer can only potentially contain a 1-bit watermark; whereas an N-bit ($N>1$) can be used for watermarking the pdf distribution of the intermediate layers, thus providing a higher capacity for IP protection. In this paper, by white-box scenario, we refer to using the whole capacity of a DL model for watermark embedding, including both hidden and output layers.

Unlike prior works, \sys{} uses the dynamic statistical properties of DL models for watermark embedding. \sys{} incorporates the watermark information within the pdf distribution of the activation maps. Note that even though the weights of a DL model are static during the inference phase, activation maps are dynamic features that are both dependent on the input keys and the DL model parameters/weights. As we illustrate in Table~\ref{tab:compuchida}, our data- and model-aware watermarking approach is significantly more robust against overwriting and pruning attacks compared with the prior-art white-box methods. None of the previous black-box deep learning WM frameworks consider overwriting attacks in their experiments. As such, no quantitative comparison is feasible in this context. Nevertheless, \sys{} performance against overwriting attacks for black-box settings is summarized in Table~\ref{tab:overwrite}.

In the rest of this section, we explicitly compare \sys{} performance against three state-of-the-art DL watermarking frameworks existing in the literature. 

\subsection{White-box Setting}
To the best of our knowledge, \cite{uchida2017embedding, nagai2018digital} are the only existing works that target watermarking hidden layers. These works use the weights of the convolution layers for the purpose of watermarking, as opposed to the activation sets used by \sys{}. As shown in~\cite{uchida2017embedding}, watermarking weights is not robust against overwriting attacks. Table~\ref{tab:compuchida} provides a side-by-side robustness comparison between our approach and these prior works for different dimensionality ratio of the attacker's WM vector to the target weights/activations. As demonstrated, \sys{}' dynamic data- and model-aware approach is significantly more robust compared to prior-art~\cite{uchida2017embedding, nagai2018digital}. As for its robustness against pruning attack, our approach is tolerant of higher pruning rates. As an example consider the CIFAR10-WRN benchmark, in which \sys{} is robust up to $80\%$ pruning rate, whereas the works in \cite{uchida2017embedding, nagai2018digital} are only robust up to $65\%$ pruning rate.

\begin{table}[ht!]
\centering
\caption{Robustness comparison against overwriting attack. The watermark information embedded by \sys{} can withstand overwriting attacks for a wide of range of $\frac{N}{M}$ ratio. In this experiment, we use the CIFAR10-WRN since this benchmark is the only model evaluated by~\cite{uchida2017embedding, nagai2018digital}.}
\label{tab:compuchida}
\resizebox{0.36\textwidth}{!}{%
\begin{tabular}{|c||c|c|}
\hline
\multirow{2}{*}{\textbf{N to M Ratio}} & \multicolumn{2}{c|}{\textbf{Bit Error Rate (BER)}} \\ \cline{2-3} 
 & \textbf{Uchida et.al}~$^{\small \cite{uchida2017embedding, nagai2018digital}}$ & \textbf{\sys{}} \\ \hline \hline
1 & 0.309 & 0 \\ 
2 & 0.41 & 0 \\ 
3 & 0.511 & 0 \\ 
4 & 0.527 & 0 \\ \hline
\end{tabular}%
}
\end{table}

\subsection{Black-box Setting}
To the best of our knowledge, there are two prior works that target watermarking the output layer for black-box scenarios~\cite{yossi, merrer2017adversarial}.  Even though the works by~\cite{yossi, merrer2017adversarial} provide a high WM detection rate (reliability), they do not address the integrity requirement, meaning that these approaches can lead to a high false positive rate in practice. For instance, the work in \cite{yossi} uses accuracy on the test set as the decision policy to detect WM information. It is well known that there is no unique solution to high-dimensional machine learning problems~\cite{choromanska2015loss, deng2014deep, goodfellow2016deep}. In other words, there are various models with even different topologies that yield approximately the same test accuracy for a particular data application. Besides high false positive rate, another drawback of using test accuracy for WM detection is the high overhead of communication and computation~\cite{yossi}; therefore, their watermarking approach incurs low efficiency. \sys{} uses a small input key size ($K=20$) to trigger the WM information, whereas a typical test set in DL problems can be two to three orders of magnitude larger.

\section{Conclusion}  
In this paper, we introduce \sys{}, the first end-to-end framework that enables reliable and robust integration of watermark information in deep neural networks for IP protection. \sys{} is applicable to both white-box and black-box model disclosure settings. It works by embedding the WM information in the probability density distribution of the activation sets corresponding to different layers of a neural network. Unlike prior DL watermarking frameworks, \sys{} is robust against overwriting attacks and satisfies the integrity criteria by minimizing the number of potential false alarms raised by the framework. We provide a comprehensive list of requirements that empowers quantitative and qualitative assessment of current and pending DL watermarking approaches. Extensive evaluations using three contemporary benchmarks corroborate the practicability and effectiveness of the \sys{} framework in the face of malicious attacks, including parameter pruning/compression, model fine-tuning, and watermark overwriting. We devise an accompanying TensorFlow-based API that can be used by data scientists and engineers for watermarking of different neural networks. Our API provides support for various DL model topologies, including (but not limited to) multi-layer perceptrons, convolution neural networks, and wide residual models.

\bibliographystyle{IEEEtran}
\bibliography{IEEEabrv,ref}
\end{document}